\documentstyle [12pt]{article}
\newcommand{\cs}[3]{{{#3} \brace {#1 #2}}}

\newcommand{\h}[1]{\mathop{\lambda}\limits_{#1}\ \!\!\!}

\newcommand{\edf}{\ {\mathop{=}\limits^{\rm def}}\ }

\begin{document}

\begin{center}
 \bf{{A NEW CLASS OF PATH EQUATIONS IN AP-GEOMETRY}}
\end{center}
\begin{center}
\bf{M.I.Wanas\footnote{Astronomy Department, Faculty of Science, Cairo University, Giza,
 Egypt\\E-mail:wanas@frcu.eun.eg} \&
 M.E. Kahil\footnote{Mathematics Department,The American University in Cairo, Cairo, Egypt\\ E-mail:kahil@aucegypt.edu}}
\end{center}
\begin{center}
\abstract{In the present work, it is shown that, the application of
the Bazanski approach to Lagrangians, written in AP-geometry and
including the basic vector of the space, gives rise to a new class
of path equations. The general equation representing this class
 contains four extra terms, whose  vanishing  reduces this
 equation to the geodesic one. If the basic vector of the
 AP-geometry is considered as playing the role of the
 electromagnetic potential, as done in a previous work, then the second term
 (of the extra terms) will represent Lorentz force while the fourth
 term gives a direct effect of the electromagnetic potential on the motion of the
 charged particle. This last term may give rise to an effect similar to the Aharanov-Bohm
 effect. It is to be considered that all extra terms will
 vanish if the space-time used is torsion-less. }
\end{center}
\section{Introduction}
In a previous attempt [1], to explore paths in AP-geometry that are
analogous to the geodesic of Riemannian geometry using the Bazanski
approach, it has been shown that there are three such paths in which
a torsion term appears. The vanishing of this term would reduce
these equations to the geodesic one. This term is found to have a
jumping coefficient from one equation to the other. One of the
authors of the present work [2] has generalized this set of
equations and suggested that the torsion term represents a type of
interaction between the torsion of the background gravitational
field and the quantum spin of the moving particle. The generalized
path equation mentioned above has been used , in its linearized
form, to interpret [3] qualitatively and quantitatively the
discrepancy in the COW-experiment. Also, the same equation has been
used [4] to explain the time delay of massless spinning particles
received from SN1987A. It is to be considered that the generalized
path equation can be used to describe trajectories of spinning
particle in any gravity theory (including General Relativity) if it
is written in the AP-geometry.

In the context of the philosophy of geometerization of physics, it
is well known that paths in an appropriate geometry are used to
represent trajectories of test particles in the background field
described completely using this geometry. For example, in
constructing his theory of general relativity (GR), Einstein has
used the paths of Riemannian geometry, the geodesic and the
null-geodesic, to represent the trajectories of scalar test
particles and of photons, respectively. Most of the achievements of
GR come from the use of these paths in studying the motion of such
particles. The path equations, in Riemannian geometry, are usually
derived by imposing an action principle on a Lagrangian function of
form (cf. [5]),
$$
L \edf g_{\mu \nu}U^{\mu}U^{\nu} \eqno{(1.1)}
$$
where $g_{\mu \nu}$ is the metric tensor and $U^{\alpha}$ is a
vector tangent to the path satisfying the conditions
$U^{\alpha}U_{\alpha}=1$ for geodesic and $U^{\alpha}U_{\alpha}=0 $
for null-geodesic. The resulting path equation can be written in the
form (cf. [6])
$$
{\frac{dU^\mu}{dS}} + \cs{\alpha}{\beta}{\mu} U^\alpha U^\beta =
0 \eqno{(1.2)}
$$
where $\cs{\alpha}{\beta}{\mu}$ is the Christoffel symbol of the
second kind. This is a well known method for deriving the path
equations in Riemannian geometry.

In the literature, there is a less known method for deriving path
equation in Riemannian Geometry. This method has been suggested by
Bazanski [7], in which an action principle is imposed on a
Lagrangian of the form,
$$
L_{B} \edf g_{\mu \nu} U^{\mu}  \frac{D \Psi^{\nu}}{DS}
\eqno{(1.3)}
$$
where $\Psi^{\nu}$ is the deviation vector and
$$
\frac{D \Psi^{\nu}}{Ds} \edf \Psi^{\nu};_{ \alpha}U^{\alpha}.
\eqno{(1.4)}
$$
The semicolon (;) denotes covariant differentiation using
Christoffel symbol. Bazanski was able{\footnote{He was also able
to derive the equation of geodesic deviation, using the same
Lagrangian (1.3), by performing variation w.r.t. the tangent
$U^{\mu}$. }} to derive the path equation (1.2) by varying the
Lagrangian (1.3) w.r.t. the deviation vector $\Psi^{\nu}$. It is
clear that the two Lagrangian (1.1) and (1.3) give identical
results, concerning paths in Riemannian geometry.

As stated above, using this approach, Wanas et al. [1] have obtained
a set of three path equations, which can be written in the compact
form,
$$
{\frac{dU^\mu}{d \tau}} + \cs{\alpha}{\beta}{\mu}~ U^\alpha U^\beta
= - ~a~ \Lambda^{~ ~ ~ ~ \mu}_{(\alpha \beta) .} ~~~U^\alpha
U^\beta, \eqno{(1.5)} $$ where $( \tau )$ is a parameter varying
along the path and $\Lambda^{\alpha}_{.\mu \nu}$ is the torsion
tensor of the AP-space. The parameter $(a)$ takes the values $0,
\frac{1}{2}, 1$ to reproduce the set of path equations mentioned
above. Wanas [2] has generalized the set (1.5) and suggested that
the torsion term in the generalized path equation may give rise an
interaction between the torsion of space-time and the quantum spin
of moving test particle. The jumping torsion coefficient, in the
path equations of the AP-geometry, may give a strong evidence that
this geometry is naturally quantized ( i.e. without performing any
known quantization schemes). Wanas and Kahil [8] have shown that the
quantum properties discovered in AP-geometry, can be found in other
non-symmetric geometries (i.e. geometries with non vanishing
torsion). It is to be considered that the path equations obtained,
in Riemannian geometry or in AP-geometry, are used to represent
trajectories of neutral test particles in a pure
 gravitational field.

 The appearance of the torsion term, with jumping coefficients, is a direct consequence of
  applying the Bazanski approach in AP-geometry. The appearance of the jumping coefficients has been
  tempting to attribute quantum properties (in the sense of Planck's quantization) to the suggested spin-torsion interaction.
 It is the aim of the present work to explore  the consequences of
 using the Bazanski approach to derive path equations in the
 presence of both gravity and electromagnetism.
 In this framework, we are going to suggest  a Lagrangian
 containing the basic vector {\footnote{ This vector has been shown
 to represent the electromagnetic generalized potential in the context of the generalized field theory
 [9].}} of the AP-geometry.
 This may throw some light on other physical interactions, if any.
 The paper is organized as follows. In section 2 we give a brief
 summary of the AP-relations needed in the present work. In section 3 we derive
 the new class of path equations. The physical meaning of the geometric terms, appeared in the new class, is given section 4. The work is discussed and concluded in section 5.
\section{Summary of Some Relations in AP-Geometry}
 AP-space is an n-dimensional manifold, each point of which is
labelled  by a set of n-independent variables $x^{\nu},( \nu =
1,2,3,...,n)$. The structure of this space is defined completely by
a set of n-contravariant vector fields $\h{i}^{\mu}$ where $i ( =
1,2,3,...,n)$ denotes the vector number, and $\mu ( = 1,2,3,...,n)$
denotes the coordinate component. The normalized cofactor
$\h{i}_{\mu}$ of the vectors $\h{i}^{\mu}$, in the determinant  $||
\h{i}^{\mu} ||$, is defined such that (cf.[10])
$$
 \h{i}^{\mu}\h{j}_{\mu} = \delta_{ij},   \eqno{(2.1)}
$$

$$
\h{i}^{\mu}\h{i}_{\nu} = \delta^{\mu}_{\nu}.  \eqno{(2.2)}
$$
Using these vectors, the following second order symmetric tensors
are defined:
$$
 g^{\mu \nu} \edf \h{i}^{\mu} \h{i}^{\nu} ,  \eqno{(2.3)}
$$
$$
g_{\mu \nu} \edf \h{i}_{\mu} \h{i}_{\nu} .  \eqno{(2.4)}
$$
Consequently,
$$
 g^{\mu \alpha}g_{\nu \alpha} = {\delta}^{\mu}_{\nu}.   \eqno{(2.5)}
$$
The tensor $g_{\mu \nu} $ can be used to play the role of the metric
tensor, of
 Riemannian space, associated with AP-space, when needed.
Consequently, using (2.3) and the derivatives (2.4), one can
define Christoffel symbols and covariant derivatives using this
symbol, in the usual manner. The following third order tensor, the
contortion tensor, can be defined as,
$$
\gamma^{\alpha}_{. \mu \nu} \edf \h{i}^{\alpha} \h{i}_{\mu ; \nu},
 \eqno{(2.6)}
$$
which is non-symmetric in its last two indices $\mu , \nu$. It can
be shown that $\gamma^{\alpha}_{. \mu \nu}$ is skew-symmetric in
its first two indices.

 It is well known that the addition of any third
order tensor to an affine connection gives another connection,
thus the object defined by,
$$
\Gamma^{\alpha}_{.\mu \nu} \edf \cs{\mu}{\nu}{\alpha} +
\gamma^{\alpha}_{.\mu \nu}, \eqno{(2.7)}
$$
is a non-symmetric connection defined in AP-geometry. So, in
addition to the covariant derivative, defined using Christoffel
symbol, one can define three more tensor derivatives using (2.7).
For example, if $A^{\mu}$ is an arbitrary contravariant vector, we
can define the following tensor derivatives for this vector,
$$
A^{\stackrel{\mu}{+}} _{.~| \nu} \edf A^{\mu} _{, \nu} +
A^{\alpha}\Gamma^{\mu}_{.\alpha \nu} \eqno{(2.8)}
$$

$$
A^{\stackrel{\mu}{-}} _{.~| \nu} \edf A^{\mu} _{, \nu} +
A^{\alpha}\tilde{\Gamma}^{\mu}_{.\alpha \nu}  \eqno{(2.9)}
$$

$$
A^{\mu}_{.~| \nu} \edf A^{\mu} _{, \nu} +
A^{\alpha}\Gamma^{\mu}_{.( \alpha \nu ) } \eqno{(2.10)}
$$
where the comma (,) denotes ordinary partial derivative, the stroke
and $(+)$ sign denotes tensor derivative using affine connection
(2.7), the stroke and the $(-)$ sign denotes the tensor derivative
using the dual connection,
$$
\tilde \Gamma^{\alpha}_{. \mu \nu} \edf \Gamma^{\alpha}_{. \nu
\mu} \eqno{(2.11)}
$$
 while  the stroke without  signs
characterizes tensor derivatives using the symmetric connection,
$$
\Gamma^{\alpha}_{. ( \mu \nu )} \edf
\frac{1}{2}(\Gamma^{\alpha}_{. \mu \nu} +  \Gamma^{\alpha}_{. \nu
\mu}), \eqno{(2.12)}
$$
or using (2.7), we can write
$$
\Gamma^{\alpha}_{. ( \mu \nu )} \edf \cs{\mu}{\nu}{\alpha} +
\frac{1}{2} \Delta^{\alpha}_{. \mu \nu} \eqno{(2.13)}
$$
where
$$
\Delta^{\alpha}_{. \mu \nu} \edf \gamma^{\alpha}_{. \mu \nu} +
\gamma^{\alpha}_{. \nu \mu} .  \eqno{(2.14)}
$$
As a consequence of using the connection (2.7), it is easy to show
that,
$$
\h{i}_{\stackrel{\mu}{+} | \nu} = 0 \eqno{(2.15)}
$$
which is usually known, in the literature, as the AP-condition.
The solution of equation (2.15) will give an alternative
definition for the non-symmetric connection,
$$
\Gamma^{\alpha}_{.  \mu \nu } \edf \h{i}^{\alpha}\h{i}_{\mu ,
\nu}. \eqno{(2.16)}
$$
Since $ \Gamma^{\alpha}_{.  \mu \nu }$ is non-symmetric, then
one can define the torsion tensor of AP-geometry as,
$$
\Lambda^{\alpha}_{.  \mu \nu }\edf \Gamma^{\alpha}_{.  \mu \nu } -
\Gamma^{\alpha}_{.  \nu \mu },  \eqno{(2.17a)}
$$
or using (2.7),

$$
\Lambda^{\alpha}_{.  \mu \nu } \edf \gamma^{\alpha}_{.  \mu \nu }
- \gamma^{\alpha}_{.  \nu \mu },  \eqno{(2.17b)}
$$
which is a skew symmetric tensor in the last two indices. As a
direct result of using (2.15) and definition (2.4), one can obtain
$$
g_{_{\stackrel{\mu}{+} \stackrel{\nu}{+} | \sigma}} =0 \eqno{(2.18)}
$$
which can be written in the form
$$ g_{\mu \nu , \sigma} = g_{\mu
\alpha} \Gamma^{\alpha}_{.\nu \sigma} + g_{\nu \alpha}
\Gamma^{\alpha}_{.\mu \sigma}. \eqno{(2.19)}
$$
Also, recall that Christoffel symbol is defined as a result of a
metricity condition,
$$
g_{\mu \nu  ;~ \sigma} = 0,  \eqno{(2.20)}
$$
which gives,
$$
g_{\mu \nu , \sigma} = g_{\mu \alpha} \cs{\nu}{\sigma}{\alpha} +
g_{\nu \alpha} \cs{\mu}{\sigma}{\alpha}. \eqno{(2.21)}
$$
In view of (2.18) and (2.20) it is clear that the operation of
raising or lowering tensor indices commutes with covariant
differentiation using Christoffel symbol and with tensor
differentiation using (2.7). Contracting (2.17b) one can obtain
the following covariant vector
$$
C_{\mu} \edf \Lambda^{\alpha}_{. \mu \alpha} = \gamma^{\alpha}_{.
\mu \alpha } \eqno{(2.22)}
$$
which is called the basic vector (cf. [11])
\section{The New Class of Path Equation}
   In the present work we are going to construct a new class of
   path equations, motivated by the following points:

1. In case of Riemannian geometry if we add a term containing  a
vector field $A^{\mu}$ to the Lagrangian (1.1) we get,
$$
L_{1} \edf g_{\mu \nu}(\hat{U}^{\mu} + \beta A^{\mu})
\hat{U}^{\nu}  \eqno{(3.1)}
$$
where $\beta$ is a converting parameter. The application of an
action principle to this Lagrangian will give the path equation (cf.
[12])
$$
\frac{d\hat{U}^\mu}{d \hat{s}} + \cs{\alpha}{\beta}{\mu}
\hat{U}^\alpha \hat{U}^\beta = - \frac{1}{2} \beta f^{\mu}_{.
\nu}\hat{U}^{\nu} , \eqno{(3.2)}
$$
where $\hat{s}$ is the evolution parameter of the path,
$\hat{U}^{\mu}$ is the tangent of the resulting path and
$$
{\it{f}}_{\mu \nu} \edf (A_{\mu, \nu} - A_{\nu, \mu} ).
\eqno{(3.3)}
$$
This is what is done to get the equation of motion of a charged
particle (mass $m$, charge $e$) in a combined gravitational and
electromagnetic field, in the context of Einstein-Maxwell's
theory. In this case, $A_{\mu}$ represents the electromagnetic
potential,
 $\beta = \frac{e}{mc}$ , ${\it{f}}_{\mu \nu}$ is the
electromagnetic field strength tensor and $c$ is the speed of
light. The term on R.H.S. of (3.2) represents Lorentz force. In
the context of the philosophy of geometerization, this approach is
subject to the objection that the vector field $A^{\mu}$ is not
part of the geometric structure.

In the context of the Generalized Field Theory (GFT), constructed in
the AP-geometry by Mikhail and Wanas [9], the vector $C_{\mu}$
defined by (2.22), represents the electromagnetic generalized
potential (using certain system of units). The skew-symmetric part
of the field equations of this theory can be written as,
$$
F_{\mu \nu} = C_{\mu , \nu} - C_{\nu , \mu} ,   \eqno{(3.4)}
$$
where $F_{\mu \nu}$ is a second order skew-symmetric tensor, defined
in the AP-geometry [9], playing the role of the electromagnetic
field strength tensor. Now $C_{\mu}$ has the physical meaning
mentioned above and is a part of the geometric structure. The
addition of this vector to the Lagrangian (1.1), in the sense given
by (3.1), will give rise to a path equation, in the AP-geometry,
similar to (3.2). This supports the physical meaning attributed to
$C_{\mu}$ and $F_{\mu \nu}$ in the framework of GFT. But this is not
our main goal. This goal will be clarified in the next
point.\\
2. Recently, some quantum (jumping) properties have been discovered
in geometries with non-vanishing torsion [1], [2] and [13]. It is
shown that these properties are closely related to affine
connections defined in the geometry. The use of Lagrangian functions
of the form given by (1.1) or (3.1), in constructing path equations,
will not give rise to such properties since it is independent of the
affine connection (cf. (1.2) and (3.2)). The Lagrangian which may
give rise to such properties should be of the Bazanski's type (1.3).
This is obvious, as it contains, in its structure, the derivative of
the deviation vector which depends on the affine connection used.
The use of such Lagrangian may give rise to quantized objects and
may
 give rise to new interactions (eg. (1.5)).
Our main goal is to modify (3.1), in AP-geometry, to include the
affine connection. This may throw some light on the quantized
quantities related to the geometric objects $C_{\mu}$ and $F_{\mu
\nu}$ and will help one to discover new interactions, if any.

In a similar way, used to modify the Lagrangian function (1.1)
 to become in the form (1.3), we are going to modify (3.1) , to be
written as,
$$
L^{*} \edf g_{\mu \nu} ( U^{* \mu} + \hat{C}^{\mu} ) \frac{D
\Psi^{\nu}}{D s^{*}} \eqno{(3.5)}
$$
where $U^{*\mu}$ is the tangent to the path and $s^{*}$ is its
evolution parameter. For dimensional consideration, concerning the
Lagrangian (3.5), we have redefined the electromagnetic potential to
be $\hat{C}^{ \mu} \edf l~ C^{\mu} $, where $l$ is a dimensional
parameter.

Now the derivative of the deviation vector in (3.5) can be
evaluated, in the AP-geometry, using one of the connections
related to (2.16). The first connection is (2.16) itself, the
second is its dual (2.11) and the third is its symmetric part
(2.13).

 In the following subsections we are going to derive the path
 equations resulting from the use of these connections.
 \subsection{The First Path Equation (using $\Gamma^{\alpha}_{. \mu \nu}$)}
 Using the non-symmetric connection given by (2.7) or (2.16), the
 Lagrangian (3.5) can be written in the form:
$$
L^{+} \edf g_{\mu \nu} (V^{\mu} + \hat{C}^{\mu})\frac{D
\xi^{\nu}}{Ds^{+}} \eqno{(3.6)}
$$
where $V^{\mu}$ is the tangent of the resulting path, $\xi^{\nu}$ is
the vector giving the deviation from this path and $s^{+}$ is the
evolution parameter along the path. The derivative of the deviation
vector is given by,
 $$
\frac{D \xi^{\nu}}{Ds^{+}} \edf \dot{\xi}^{\nu} +
\xi^{\alpha}\Gamma^{\nu}_{.\alpha \beta} V^{\beta} \eqno{(3.7)}
 $$
where, $$\dot{\xi^{\nu}} \edf \frac{d \xi^{\nu} }{ds^{+}}.
\eqno{(3.8)} $$
 Now, we have
 $$
\frac{\partial L^{+}}{\partial\dot{\xi}^{\sigma}} = g_{\mu
\sigma}(V^{\mu}+ \hat C^{ \mu}).
 $$
Then, making use of (2.19) we can write
$$
\frac{d}{ds^{+}}\frac{\partial L^{+}}{\partial\dot{\xi}^{\sigma}}
= g_{\mu \sigma}\frac{d}{ds^{+}}(V^{\mu} +\hat{C}^{\mu}) + [
g_{\lambda\sigma}\Gamma^{\lambda}_{.\mu \rho} +
g_{\mu\lambda}\Gamma^{\lambda}{. \sigma \rho}]( V^{\mu} +
\hat{C}^{\mu})V^{\rho}. \eqno{(3.9)}
$$
Also from (3.6) we can write,
$$
\frac{ \partial L^{+}}{\partial{\xi}^{\sigma}} = g_{\mu
\nu}(V^{\mu} + \hat{C}^{\mu})\Gamma^{\nu}_{. \sigma
\beta}V^{\beta}. \eqno{(3.10)}
$$
Substituting (3.9) and (3.10) into the Euler-Lagrange equation,
$$
\frac{d}{ds^{+}}\frac{\partial L^{+}}{\partial
\dot{\xi}^{\sigma}}~ -~ \frac{\partial
L^{+}}{\partial{\xi}^{\sigma}} =~ 0 ,
$$
we get, after necessary reductions, the path equation
$$
{\frac{dV^\mu}{ds^+}} + \cs{\alpha}{\beta}{\mu} V^\alpha V^\beta =
- \Lambda^{~ ~ ~ ~ \mu}_{(\alpha \beta) .} ~~V^\alpha V^\beta -
\hat{F}^{\mu}_{. \nu}V^{\nu} - g^{\mu
\delta}\hat{C}_{_{\stackrel{\nu}{+}} | \delta}V^{\nu}- \Lambda^{~
~ ~ ~ \mu}_{\alpha \beta .} ~~\hat{C}_{\alpha} V^\beta   ,
\eqno{(3.11)}
$$
where
$$
\hat{F}_{ \mu \nu} \edf  \hat{C}_{\mu, \nu} - \hat{C}_{\mu , \nu}.
$$
\subsection{The Second Path Equation (using $\Gamma^{\alpha}_{.( \mu \nu )}$)}
We use in this subsection the symmetric connection given by (2.13)
to evaluate the Lagrangian (3.5), which can be written as,
$$
L^{o} \edf g_{\mu \nu}(W^{\mu}+ \hat{C}^{\mu}) \frac{D \zeta^{
\nu}}{D s^{o}} \eqno{(3.12)}
$$
where $W^{\mu}$ is the tangent to the resulting path,
$\zeta^{\nu}$ is the deviation vector and $s^{o}$ is the parameter
varying along the path. Similar to subsection 3.1 we can write the
definitions
$$
\frac{D \zeta^{\nu}}{Ds^{o}} \edf  \dot{\zeta}^{\nu} +
\zeta^{\alpha}\Gamma^{\nu}_{.(\alpha \beta)} W^{\beta}
\eqno{(3.13)}
$$
and
$$\dot{\zeta}^{\nu} \edf \frac{d \zeta^{\nu}}{ds^{o}}.$$
Evaluating the variational derivatives of the Lagrangian (3.12) and
substituting in the Euler-Lagrange equation, as done in section 3.1,
we get after some, relatively long but straightforward,
 calculations the second path equation
$$
{\frac{dW^\mu}{ds^o}} + \cs{\alpha}{\beta}{\mu} W^\alpha W^\beta =
-\frac{1}{2} \Lambda^{~ ~ ~ ~ \mu}_{(\alpha \beta) .} ~~W^\alpha
W^\beta - \hat{F}^{\mu}_{. \nu}W^{\nu} - g^{\mu \delta}
\hat{C}_{_{\stackrel{\nu}{+}} | \delta}W^{\nu} - \frac{1}{2}
\Lambda^{~ ~ ~ ~ \mu}_{\alpha \beta .} ~~\hat{C}^{\alpha} W^\beta
. \eqno{(3.14)}
$$
\subsection{The Third Path Equation (using $\tilde\Gamma^{\alpha}_{.\mu \nu}$)}
For this path equation, the Lagrangian (3.5) can be written in the
form,
$$
L^{-} \edf g_{\mu \nu} (J^{\mu} + \hat{C}^{\mu}) \frac{D
\eta^{\nu}}{Ds^{-}}   \eqno{(3.15)}
$$
where $J^{\mu}$ is the tangent to the path, $\eta^{\nu}$ is the
deviation vector and $s^{-}$ is the evolution parameter
characterizing the path. Similar to the definitions given in the
previous subsections we write,
$$
\frac{D \eta^{\nu}}{Ds^{-}} \edf \dot{\eta}^{\nu} +
\eta^{\alpha}\tilde{\Gamma}^{\nu}_{. \alpha \beta} J^{\beta} ,
\eqno{(3.16)}
$$
where,
$$
\dot{\eta}^{\nu} \edf \frac{d \eta}{ds^{-}}. \eqno{(3.17)}
$$
Performing necessary variational calculations, as done in the
previous subsections and substituting in the Euler-Lagrange
equation we get, after some rearrangements, the third equation
$$
{\frac{dJ^\mu}{dS^-}} + \cs{\alpha}{\beta}{\mu} J^\alpha J^\beta =
- \hat{F}^{ \mu}_{. \nu}J^{\nu} - g^{\mu
\delta}\hat{C}_{_{\stackrel{\nu}{+}} | \delta}J^{\nu}.
\eqno{(3.18)}
$$
\section{Physical Meaning of the Geometric Terms}
The set of path equations (3.11), (3.14) and (3.18) comprises a new
class of path equations which can be written in the general form
$$
{\frac{dZ^\mu}{d\tau}} + a_{1}~ \cs{\alpha}{\beta}{\mu} Z^\alpha
Z^\beta = -~a_{2}~ \Lambda^{~ ~ ~ ~ \mu}_{(\alpha \beta) .}
~~Z^\alpha Z^\beta -~a_{3}~ \hat{F}^{\mu}_{. \nu}Z^{\nu} -~a_{4}~
g^{\mu \delta}\hat{C}_{_{\stackrel{\nu}{+}} | \delta}Z^{\nu} -~
a_{5}~ \Lambda^{~ ~ ~ ~ \mu}_{\alpha \beta .} ~~\hat{C}^{\alpha}
Z^\beta , \eqno{(4.1)}
$$
where $Z^{\mu}$ is the tangent of the resulting path , $\tau$ is
the evolution parameter of the path, $a_{1}, a_{2}, a_{3}, a_{4},
$ and $a_{5}$ are numerical parameters whose values are listed in
the following table.
\begin{center}
 {Table 1: Values of the parameters of (4.1) corresponding to connection used}
\end{center}
\begin{center}
\begin{tabular}{|c|c|c|c|c|c|c|} \hline
 Connection used&$a_{1}$&$a_{2}$& $a_{3}$& $a_{4}$ & $a_{5}$ & Equation  \\
 \hline
& & & & & & \\
  ${\Gamma}^{\alpha}_{.~\mu \nu} $ & 1& 1& 1& 1 & 1& (3.11)       \\
& & & & & & \\
\hline
& & & & & & \\
   ${\Gamma}^{\alpha}_{.~ (\mu \nu )}$ &1 &  $\frac{1}{2}$ & 1 & 1 & $\frac{1}{2}$ & (3.14)    \\
& & & & & & \\
   \hline
& & & & & & \\
  $\tilde{\Gamma}^{\alpha}_{.~\mu \nu } $ & 1&  0 & 1 & 1& 0 & (3.18)  \\
& & & & & & \\
 \hline
\end{tabular}
\end{center}
In the context of the scheme of geometerization, if the general
equation (4.1) is used to study trajectories  of charged test
particles, we can attribute the following physical meanings to
some of its terms: \\
 1- The term whose coefficient is $(a_{1})$,
$\cs{\alpha}{\nu}{\mu} Z^{\alpha }Z^{\nu}$,
represents , as usual, the effects of
 gravitation on  the motion of the particle. \\
2- The term whose coefficient is $(a_{2})$, $\Lambda_{( \alpha \nu)
.}^{~~~\mu} Z^{\alpha} Z^{\nu}$, is suggested [2]
 to represent a type of interaction between the torsion of space-time and quantum spin
  of the moving particle. This interaction will affect the motion of the particle.
This term is quantized as clear from the values of $(a_{2})$ in
Table 1. It is shown that
 there are some experimental [3] and observational [4]
 evidences for the existence of this interaction. \\
3- The term whose coefficient is $(a_{3})$, $ \hat{F}^{\mu}_{.
\beta}Z^{\beta} $,
 represents the effect of the electromagnetic field on the motion of a charged
 particle,the
  Lorentz force, as obvious from the comparison with the
  R.H.S. of (3.2). The coefficient of this term, as clear from Table  1,
   does not vary from equation to another. So, this effect is not quantized. \\
4- The term whose coefficient is $(a_{5})$, $ \Lambda_{ \alpha \nu
.}^{~~~\mu} \hat{C}^{\alpha} Z^{\nu}$ represents an interaction
 between the  electromagnetic potential and the torsion of space-time.
 The term is already quantized, as clear from Table 1. This term
 represents a direct effect of the electromagnetic potential on
 the motion of a charged particle similar to the Aharonov-Bohm
 (AB) effect [14], [15], with one main difference, that is the influence of
 the torsion on this term. This will be discussed in the
 following section.
\section{Discussion and Concluding Remarks}
In the present work, we derived a new class of path equation, in
AP-geometry, using Bazanski method. All terms appearing in this
class are parts of the geometric structure used. The general
equation, representing this class can be used, qualitatively, to
explore different interactions affecting trajectories of charged
particles in a combined gravitational and electromagnetic field.

If the terms representing the electromagnetic effects on the
trajectory, $\hat{F}_{\mu \nu}$ and $\hat{C}_{\alpha}$, are switched
off, then the new class will reduce to the old set of path equations
[1] with spin-torsion term. If further, we neglect this term we get
the ordinary geodesic equation.

Two of the terms, on the R.H.S. of the general equation (4.1), are
naturally quantized in Planck's sense ( terms with jumping
coefficients). One of these terms is the spin-torsion term with the
coefficient $(a_{2})$ and the other is the term giving rise to
AB-type effect (the term with coefficient $(a_{5})$, see Table 1 ).
All other terms, including the Lorentz force term, of this equation
are not quantized, in the above mentioned sense. It is clear from
this equation that the appearance of the quantum properties in this
equation is closely connected to the explicit appearance of the
torsion in the terms concerned. Since Riemannian geometry is a
torsion free geometry, such quantum properties did not appear in any
field theory constructed in this geometry.

We would like to point out that we are not claiming that we are
doing Quantum Mechanics or Quantum Field Theories. Actually, we are
dealing with a property, in AP-geometry that some terms of the
equations have jumping coefficients, which reflects some quantum
features in the sense of Planck's quantization.
  It is well known, in the literature, that the AB-effect is a quantum phenomenon which
is impossible to be accounted for using classical electrodynamics.
The scheme followed in present work is a pure geometric scheme,
which is considered by many authors as a classical scheme. Now, if
the term, whose coefficient is $a_{5}$, is interpreted to give rise
to AB-effect, then one can draw the following conclusion : {\it
{Either the AB-effect is a classical phenomenon and can be accounted
for using a classical scheme, or the type of geometry used, the
AP-geometry, admits some quantum properties}}.

\section*{References}
{[1]} Wanas, M.I., Melek, M. and Kahil, M.E.(1995) Astrophys. Space
Sci.,{\bf{228}}, 273. ;

gr-qc/0207113. \\ \\ {[2]} Wanas, M.I.(1998) Astrophys. Space Sci.,
{\bf{258}}, 237 ;

 gr-qc/9904019. \\ \\
 {[3]} Wanas, M.I., Melek, M. and Kahil, M.E. (2000) Grav.
Cosmol., {\bf{6} }, 319. \\ \\
{[4]} Wanas, M.I., Melek, M. and Kahil, M.E. (2002) Proc. MG IX,
part B, p.1100, Eds.

V.G. Gurzadyan et al. (World Scientific Pub.); gr-qc/0306086. \\ \\
{[5]} Adler, R.  Bazin, M. and Schiffer, M. (1975)
"{\it{Introduction to General Relativity}}",

 McGraw Hill.  \\ \\
{[6]} Eisenhart, L.P. (1926) "{\it{Riemannian Geometry}}", Princeton
Univ. Press.\\ \\
{[7]} Bazanski, S.I. (1989) J. Math. Phys.,
{\bf{30}}, 1018.
\\ \\
{[8]} Wanas, M.I. and Kahil, M.E.(1999) Gen. Rel. Grav., {\bf{31}},
1921. ;

 gr-qc/9912007 \\ \\
{[9]} Mikhail, F.I. and Wanas, M.I. (1977) Proc. Roy. Soc. Lond.
{\bf{A 356}}, 471. \\ \\ {[10]} Wanas, M.I. (2001) Stud. Cercet. \c
Stiin\c t. Ser. Mat. Univ. Bac\u au {\bf{10}}, 297;

 gr-qc/0209050 \\ \\ {[11]} Wanas, M.I. (2000) Turk. J. Phys., {\bf{24}}, 473 ;

gr-qc/0010099. \\ \\
{[12]} Straumann, N. (1984) "{\it{General
Relativity and Relativistic Astrophysics}}",

 Springer-Verlag. \\ \\
{[13]} Wanas, M.I. (2003) Algebras, Groups and Geometries, {\bf 20},
345a.   \\ \\
{[14]} Aharanov, Y. and Bohm, D. (1959) Phys. Rev. {\bf{115}}, 485.
\\ \\ {[15]} Peshkin, M. (1981) Physics Reports, {\bf{6}}, 375.
\end{document}